%% file: main.tex
\documentclass[runningheads]{llncs}
\usepackage{graphicx}
\usepackage{subfiles}
\usepackage{hyperref}
\usepackage[square, numbers]{natbib}
\usepackage{dirtytalk}
\usepackage{wrapfig}

\begin{document}
\title{Ranking Clarifying Questions Based on Predicted User Engagement}
%
%
\author{Tom Lotze \and
Stefan Klut \and Mohammad Aliannejadi \and Evangelos Kanoulas
%
%
\institute{University of Amsterdam}
\email{Tom.Lotze@gmail.com, stefanklut@outlook.com, \{M.Aliannejadi, E.Kanoulas\}@uva.nl}}
\maketitle              
\begin{abstract}
To improve online search results, clarification questions can be used to elucidate the information need of the user. This research aims to predict the user engagement with the clarification pane as an indicator of relevance based on the lexical information: query, question, and answers. Subsequently, the predicted user engagement can be used as a feature to rank the clarification panes.
   
Regression and classification are applied for predicting user engagement and compared to naive heuristic baselines (e.g. mean) on the new MIMICS dataset \cite{Zamani_MIMICS}. An ablation study is carried out using a RankNet model to determine whether the predicted user engagement improves clarification pane ranking performance.
The prediction models were able to improve significantly upon the naive baselines, and the predicted user engagement feature significantly improved the RankNet results in terms of NDCG and MRR. This research demonstrates the potential for ranking clarification panes based on lexical information only and can serve as a first neural baseline for future research to improve on. The code is available online\footnote{\url{https://github.com/Tom-Lotze/IR2_5}}.

\end{abstract}
\section{Introduction}\label{sect:Introduction}
\subfile{Sections/1-Introduction}

\section{Related Work}\label{sect:Related}

\subfile{Sections/2-RelatedWork}
\section{Methods}\label{sect:Methods}
\subfile{Sections/3-Methods}

\section{Results}\label{sect:Results}

\subfile{Sections/4-Results}

\section{Conclusion \& Limitations}\label{sect:Conclusion}
\subfile{Sections/5-Conclusion}


%
%
%
\bibliographystyle{splncs04}
\bibliography{main}



\end{document}

%% file: Sections/1-Introduction.tex

Search clarification is an upcoming topic in the field of Information Retrieval and, more specifically, conversational search. The increasing use of devices with a small screen - or speech-only interface - further increases the need for correctly predicting the intent of the query, as showing multiple documents might be impractical or impossible. Query clarification, in which a clarifying question is asked to the user about intent and information need, is one way to ensure more relevant results.

\begin{figure}[ht]
    \centering
    \includegraphics[width=0.48\linewidth]{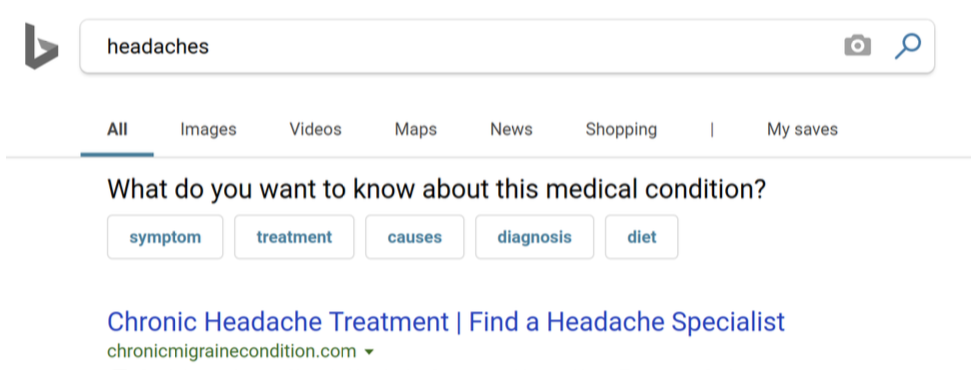}
    \includegraphics[width=0.48\linewidth]{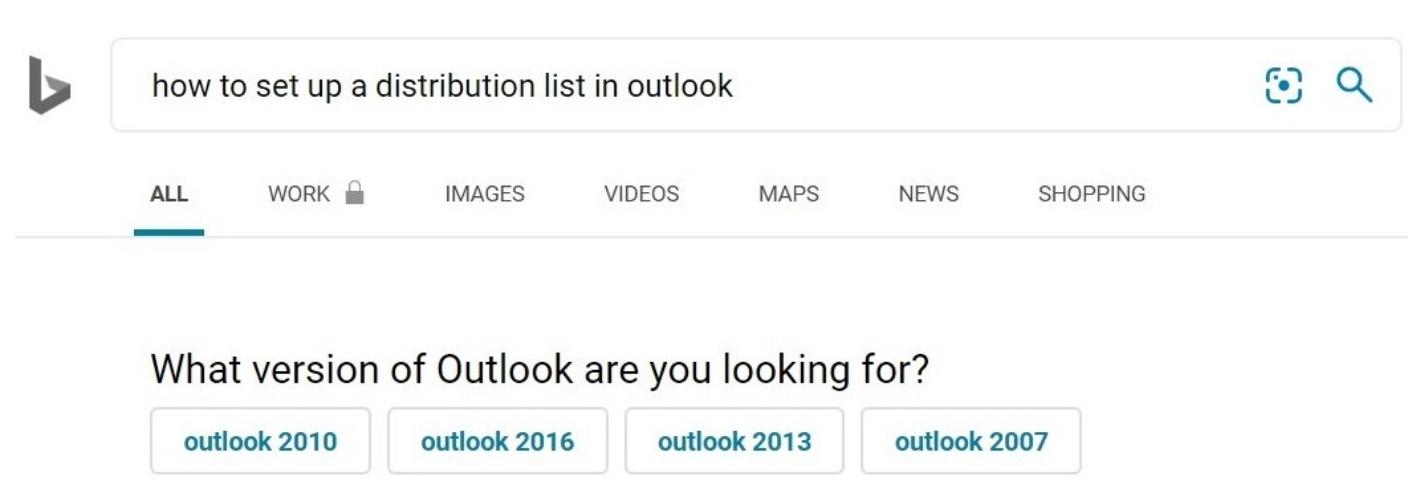}
    \caption{Examples of the CP, taken from \cite{Zamani_generation, Zamani_MIMICS}}
    \label{fig:clarification_pane}
\end{figure}

Microsoft has introduced such a query clarification method for its search engine \textit{Bing}. This method uses generated questions and answers based on reformulation data \cite{Zamani_generation}. The questions are presented in the form of Clarification panes (CPs): a question with $2$ to $5$ corresponding answers to refine the query. Examples of queries with their corresponding CP are shown in Figure \ref{fig:clarification_pane}. A dataset containing CPs and corresponding user interaction was recently released by \cite{Zamani_MIMICS} and is called MIMICS. It is one of the first large datasets on conversational search containing user interaction data.

Identifying which CPs will be engaged with may be used to improve user experience by presenting them with better questions. If we assume that predicted user engagement (PUE) with these CPs can be used as a proxy for the relevance of the question, we can formulate selecting the best CP as a learning to rank problem, based on PUE.

However, actual user engagement (UE) is not always available, and ranking without the user interactions is a very difficult problem. A more realistic and useful scenario would be to access the lexical information, such as query, question, and answer options. By predicting the UE using only these lexical features, we can try to rank the CPs without actual access to the user interactions. In other words, the goal of this paper is to propose a method that takes only the lexical features of a query-question pair, which it uses to predict the UE on this CP. This PUE will then be used as a feature in the second and final step of the pipeline: Learning to rank. More specifically, this goal can be expressed in the following three research questions:\\

\noindent \textbf{RQ1}: Can we accurately model user engagement on clarification panes as a regression and classification problem, and what is the difference in performance, also compared to naive baselines?\\
\textbf{RQ2}: What is the difference in performance between BERT and TFIDF representations of the lexical information on predicting user engagement through either classification or regression?\\
\textbf{RQ3}: Can we improve in ranking the clarification panes by using predicted user engagement as a feature in Learning to Rank?\\

To the best of the authors' knowledge, this is the first attempt to predict the UE based solely on the query, questions, and answers on this new dataset and use it to rank the CPs. We aim to construct the first baseline on this task without using any real user interaction data. Ultimately, this can be useful for improving the user experience in an online search. CPs with high PUE may be shown in favor of panes with low UE and better guide the user in its search, yielding more user satisfaction \cite{Aliannejadi_questions_in_IR,DBLP:journals/corr/abs-2009-11352}.


%% file: Sections/2-RelatedWork.tex

\subsection{Conversational Search}
Conversational search is a domain in Information Retrieval, describing a specific form of human-computer interaction. The defining feature of this field is that the device responds to the human, creating a two-way, mixed-initiative interaction system \cite{Radlinski_framework, Kiesel_voice_query}, such as chat-bots or speaking digital assistants. These examples illustrate the two main forms of conversational search: Spoken and textual interfaces \cite{Kiesel_voice_query}. Nevertheless, both variants only allow for returning a small number of documents (often just one) \cite{Aliannejadi_questions_in_IR, Zamani_analysis}, a crucial factor in the need for search clarification.

\subsection{Search clarification: Clarifying Questions}
The importance of clarification in search starts at the user queries. Queries are often short, possibly ambiguous, and faceted, and the intent and information needs of the user might not be clear \cite{Zamani_generation, Zamani_analysis, Krasakis_document_ranking}. In other words, the users \say{often fail to compose a single query that entails all their information needs} \cite{Aliannejadi_questions_in_IR}. An example could be the query \say{\textit{Jaguar}}, which could refer to either the animal or the car brand.

One solution to this would be search result diversification \cite{Santos_query_reformulation, Drosou_search_diversification, Zamani_generation}, which aims to cover most needs and facets of a single query in the returned documents. However, this method's underlying assumption - the possibility to present a ranked list to the user - does not hold for small-screen interfaces, speech devices, or question-answering systems \cite{Zamani_analysis, Aliannejadi_questions_in_IR}. An alternative approach is to inquire the user about the intent of their query by means of one or more clarifying questions \cite{Zamani_generation, Hashemi_neural_interaction_model}. These questions can either have closed form answers, which also have to be generated, or it can be open questions, as done in \cite{Krasakis_document_ranking}. Our research's scope is limited to fixed candidate answers, but posing open clarification questions has already been shown to improve search results significantly \cite{Hashemi_neural_interaction_model}. For the methodology on generating clarifying questions and their answers, we refer the reader to \cite{Zamani_generation}.

With experts assessing the quality of the CPs as good in general \cite{Zamani_generation}, there is one crucial question left: Does the user experience the CPs as helpful for finding relevant documents? Without the use of search clarification, (ambiguous) queries may lead to iterations of query reformulation or scanning multiple result pages, reducing user satisfaction \cite{Aliannejadi_questions_in_IR}. On the other hand, \cite{Kiesel_voice_query} showed that users are not dissatisfied when prompted for voice query clarification and might even like it. Furthermore, the click-through rate relatively improved by a significant $48\%$ compared to query suggestion \cite{Zamani_generation}.

\subsubsection{User engagement}
To estimate the usefulness of different clarification questions, it is important to measure the UE with the CPs. However, it is not possible to observe the actual engagement with the clarification questions. Merely we can observe user interactions with the CP. Many types of interactions can be used to describe the UE, for example, the click-through rate or the dwell time \cite{Lehmann_user_engagement_measures, OBrien_Theoretical_UE}. 

Research into UE in real-world applications has already been done. \cite{Mitra_user_interaction_autocomplete} showed that for query autocompletion, the suggested completion position has a big impact on user interaction. The top-ranked completion was chosen more than twice as much as the second-highest ranked completion. \cite{Mitra_user_interaction_autocomplete} also showed that the class of question (e.g., finance or celebrity) influences the UE with the autocomplete results. For autocompletion, the number of characters typed already was also of influence. However, unlike autocompletion, for clarification questions, the query is already typed out in full. 

A highly relevant work on predicting user engagement was published recently, making use of the same MIMICS dataset \cite{sekulic2021user}. In this paper, the authors try to predict the user engagement on clarification panes using a Transformer-based model \cite{sekulic2021user}. Their objective is similar to ours, namely to use the user engagement as a proxy for relevance. However, the clarification panes are not reranked based on this. Instead, the focus lies on incorporating parts of the SERP into the model, to improve the accuracy in predicting user engagement.


\subsubsection{Ranking of CPs}
In \cite{Zamani_analysis}, a first attempt was also made to re-rank the questions to suit the users' needs better, but with real user interaction, which is not always available. Moreover, \cite{Rosset_suggesting_useful_questions} rank follow-up questions based on their usefulness to the user. However, this usefulness is based on the assumption that the user's information needs were satisfied using their original query. The follow-up questions are based on different information needs a user might have after finding the answer to the original query, which may help to improve the \say{People Also Ask} section in modern search engines. In our research, the clarification questions for a given query specifically help elucidate the current information need.

To predict whether a CP will be relevant to the user, it is important to predict UE on clarifying questions accurately. Predicting UE has been done in the past in different contexts. For example, \cite{Straton_facebook_prediction} used (deep) neural nets to predict UE on Facebook posts based on post characteristics (e.g. time, post type or country). More recently, \cite{Risch_comment} also used neural networks to predict UE with comments on news articles. To the best of the authors' knowledge, predicting UE for CPs has not been done before. Unlike previous UE prediction attempts, the research in this paper takes (predicted) user engagement as a measure of relevance instead of a measure of popularity. In this paper we will examine whether this assumption holds and use the PUE as a feature for ranking the CPs.

\subsection{Datasets}
The main reason no prior research has been done on predicting UE on clarification questions was the lack of a sufficiently large data set. Until recently, the only dataset that existed specifically on search clarification was Qulac \cite{Aliannejadi_questions_in_IR}. However, this dataset consists of only $200$ queries and is therefore not suited to train (large) neural networks. The questions and answers in this dataset were also not obtained from real users but rather through crowd-sourcing, which might affect the quality of the data. 

The publication of the MIMICS dataset by \cite{Zamani_MIMICS} changed this, as the first part of this dataset (MIMICS-Click) consists of approximately $400k$ unique queries, with the associated CP and user interaction signals. MIMICS-Click uses clicks as a measurement for user interaction. Receiving no click on a CP would be a UE of $0$, and all other clicks are binned into engagement levels from $1$ to $10$, using the equal-depth method \cite{Zamani_MIMICS}. The equal-depth binning implies that all nonzero bins contain approximately the same number of samples. The data was taken from real user interaction with the \textit{Bing} search engine. All queries in the dataset have a corresponding CP and are completely in English. Another part of the dataset (MIMICS-ClickExplore) consists of approximately $60k$ unique queries, where each query has multiple different CPs. Using this dataset, we can explore and predict UE on the various CPs corresponding to one query and rank those.

%% file: Sections/3-Methods.tex
In this section, the dataset will first be explored, and various pre-processing options are explained for the prediction part of the pipeline (regression / classification) and the ranker part. Next, the models will be described, including relevant design choices, hyperparameters, and training schemes. Also, the evaluation metrics and the experimental design are laid out. Finally, the baselines for the experiments and significance testing are clarified.

\subsection{Dataset}\label{sec:data}
\subsubsection{Dataset: Regression \& Classification}
For the regression and classification part of the pipeline, the MIMICS-Click dataset will be used \cite{Zamani_MIMICS}. We will use the user-submitted query for the regression and classification models, the generated clarifying question, and the candidate answers. The input of the prediction model is a vectorized representation of these features, using either a BERT embedding or TFIDF representation. For each data instance, an impression level is available (\say{\textit{low}}, \say{\textit{medium}}, \say{\textit{high}}), indicating the level of exposure to users. The target is the UE, represented as an integer value in range $[0, 11]$. The label distribution is severely imbalanced, with $83\%$ of engagement labels being $0$ (i.e., no engagement at all), which is solved by balancing the dataset.

The following Boolean pre-processing options were designed and evaluated using regression and classification models: $1$) \textbf{Balance distribution}: When true, the frequency of engagement level $0$ will be reduced to the median frequency of all labels. In all prediction experiments, the dataset is balanced. $2$) \textbf{Impression level filter}: When set to true, data instances with a \say{\textit{low}} impression-level are removed from the dataset to reduce noise. $3$) \textbf{Reduced classes}: if set to true, the engagement labels are mapped to a binary distribution: No exposure vs. any exposure. $4$) \textbf{Vectorizer}: Can be either \say{BERT} or \say{TFIDF}. If set to \say{BERT}, a sentence embedding is obtained for each remaining data instance using a fixed \texttt{DistilBERT} model, which is not further optimized \cite{Reimers_sentence-bert}. We concatenate these $7$ embeddings (query, question, $5$ answers) of size $768$ to obtain a 1D vector (of size $5376$). If set to \say{TFIDF}, the TFIDF vectorizer from \texttt{scikit-learn} is used to obtain a 1D representation per data instance, with a fixed vocabulary size of $30k$. The resulting dataset is randomly split in a train-, validation- and test set, with respective splits being $70\%$, $15\%$, $15\%$. Each data split is shuffled and split up into batches of size $64$ (the last incomplete batch is dropped). The number of epochs is set to $40$ for both classification and regression, as convergence analysis showed all used parameter settings had converged by then.

\subsubsection{Dataset: Ranker}
For the ranker, the MIMICS-ClickExplore dataset is used \cite{Zamani_MIMICS}. This dataset contains $64,007$ unique user queries, each with at least two corresponding CPs that have been shown to users. The maximum number of questions that are associated with a single query is $89$. There is a total of $168,921$ query-question pairs, which means there are $2.64$ clarifying questions per query on average. The engagement levels distribution in the MIMICS-ClickExplore dataset, while not as severe as the MIMICS-Click dataset, is still unbalanced. There are $89,441$ query-question pairs with a non-zero engagement level, which means that approximately $53\%$ of all pairs has an engagement level of $0$. However, in this case, we keep all the data instances given the low number of CPs to rank per query. 

With the low number of average questions per query, negative samples were introduced in the dataset. These negative samples are random CPs corresponding to the different queries. For each query, $10$ negative samples are added to the questions. All negative samples receive a relevance label of $0$, the positive samples with maximum predicted engagement receive a label of $2$, and all other positive samples receive a label of $1$. The pre-processed dataset was split into a training, validation, and test set, with respective splits of $70\%$, $15\%$, and $15\%$. However, for this dataset, the splits were not done on query-question pairs but solely on the queries to ensure that the same query's questions correspond to the same query end up in the same split.

\subsection{Engagement Prediction}
Given the dataset and possible pre-processing options, this research's first goal is to produce an accurate regression and classification model for engagement prediction on the CPs. The regression and classification models have an identical architecture, except for the last layer (mapping to a scalar or to the number of classes).
We set an evaluation frequency on the validation set of $100$ batches for both regression and classification. The model parameters are saved at the point in training where the validation loss is minimal. After convergence ($40$ epochs), the optimal model is loaded from memory and evaluated on the test set.


\subsubsection{Model parameters}\label{sec:parameters}
A \texttt{PyTorch} Multi-Layer Perceptron (MLP) is used for the regression \cite{Paszke_pytorch}. The Mean-Square Error (MSE) loss is used for the regressor, and the Cross-Entropy (CE) loss in the classifier. The optimal layer configuration and other hyperparameter settings are found by extensive hyperparameter tuning. The optimal parameters are defined by the MSE-loss on the test set. Due to computational constraints, and given the similarity of the classification and regression model, the optimal parameters found for regression are also used for the classification model. The following optimal values were found: Both the regression and classification model have two hidden layers of size $300$ and $32$ and make use of a \texttt{LeakyReLU} activation function with a negative slope of $0.02$, as well as batch normalization to ensure stable training. The \texttt{AmsGRAD} optimizer is used (SGD and Adam were also tested, with momentum), with a learning rate of $0.001$, and no weight decay.


\subsection{CP Ranking}
The second goal of this research is determining whether the PUE found using the previously trained regression or classification model may be used to improve the quality of a ranking algorithm. To see if there is any impact an ablation study is performed, where one RankNet model uses the PUE as one of the input features and another identical RankNet model does not. Besides the PUE, Ranknet takes as a input the number of characters in the query and question, the number of answers, and the average number of characters in the answers (all normalized). Next, the number of pairwise errors found in the ranking is minimized \cite{Burges_baselines}. If improvements are found for RankNet, it is plausible they will also improve more sophisticated models based on RankNet, such as LambdaRank. The sped-up version of RankNet is used so all gradients for all pairwise comparisons can be used at once to update the weights of the model.

Both the RankNet with and without the predictions are trained for $5$ epochs. The performance of the model is evaluated using the Normalized Discounted Cumulative Gain (NDCG) and Mean Reciprocal Rank (MRR) measures. NDCG is a metric which compares the returned ranking with the ideal possible ranking to achieve a score that shows how good a ranking is regardless of the number of documents. A higher NDCG means that the ranker performs better. The MRR is an indicator for the reciprocal rank of the first relevant document. A higher MRR means that the first relevant document is found higher in the ranking. After training the model is evaluated in terms of NDCG and MRR on the held-out test set. A perfect model would score $1.0$ on both NCDG and MRR.


\subsection{Baselines}
Simple baselines are constructed for fair evaluation of our prediction models. For the various data pre-processing settings, the mean, median and mode of the engagement labels are computed and the corresponding test loss is compared to the performance of our models. Significance testing is done for every baseline, and between the regression and the classifier (rounding the regression predictions). A paired two-sided t-test was used. The null hypothesis (the two samples have identical average (expected) values) will be rejected if the p-value $<0.001$. 

The ablation results for the ranker are compared in terms of NDCG and MRR, using the same significance test for both the NDCG and MRR separately. For this comparison we will use a larger p-value cut-off of $0.05$, as we are not comparing to a naive heuristic baseline.

%% file: Sections/4-Results.tex
\begin{table*}[htb]
\centering
\caption{Test loss for all data settings and models, compared to baselines. Columns represent pre-processing steps described in section \ref{sec:data}: Vector representation, Impression level filter (\texttt{T}rue/\texttt{F}alse). 11 classes are used. Best value per method per column is indicated in bold. An * indicates significant difference of the baseline compared to our method.}
\begin{tabular}{|l||c|c|c|c|}
\hline
\begin{tabular}[c]{@{}l@{}}Model / Baseline\end{tabular} &
  TFIDF, T &
  TFIDF, F &
  BERT, T &
  BERT, F \\ \hline\hline
\begin{tabular}[c]{@{}l@{}}Regression \\ (\textbf{MSE-loss})\end{tabular} 
        & \textbf{5.553}  & \textbf{9.518}  & \textbf{5.351} & \textbf{9.058} \\ \hline
Mean    & 5.985* & 9.597* & 5.985*  &  9.597*  \\ \hline
Median  & 6.030* & 9.607* & 6.030* &  9.607*  \\ \hline
Mode    & 6.030* & 19.200* & 6.030* &  19.200*  \\ \hline\hline
\begin{tabular}[c]{@{}l@{}}Classification \\ (\textbf{CE-loss / accuracy})\end{tabular} 
        & \textbf{2.10/18.7}  & \textbf{2.353/14.3}  & \textbf{2.081/19.6}  & \textbf{2.34/14.2} \\ \hline
Median = Mode  & 2.375/16.9* & 2.440/10.6* & 2.375/16.9 & 2.440/10.6* \\ \hline
Random model & 2.375/16.9* & 2.440/10.6* & 2.375/16.9  & 2.440/10.6* \\ \hline
\end{tabular}
\label{tab:results}
\end{table*}

In this section, the results from the experiments described in section \ref{sect:Methods} are shown and analyzed. First, the results for the prediction part are analyzed for both the regression and classification approaches. Subsequently, the results for the ranker ablation study are presented and analyzed.

\subsection{Engagement Prediction}
In Table \ref{tab:results} an overview is given of the performance of both the regression and classification loss for the pre-processing data options (see section \ref{sec:data}), as well as the accuracy for the classification approach. The same metrics are shown for the simple baselines mean, median, and mode. Please note that the baselines are independent of the vectorizer. All zero-labels are balanced, as explained in the methods section, but due to dropping the last incomplete batch, the final label distribution of the test set might not be fully uniform.

It can be observed from Table \ref{tab:results} that our regression and classification models consistently outperform the simple baselines (RQ1). Significance testing showed that almost all configurations are significantly better than the baselines. Note that the mode can vary drastically depending on the distribution of the data (in some cases, the mode might not be close to the mean or median). 

Moreover, the table shows that BERT embeddings consistently outperform the TFIDF features (RQ2), regardless of the other data settings. Furthermore, as expected, the models perform better on only two classes (results not shown). A significance test was also performed on the classifier vs. the regression for all $8$ settings by rounding the regression predictions to the nearest integer. For all settings, the predictions on the test set are significantly different.

\subsubsection{Regression}
We observe substantial overfitting for every model, despite the relatively small architecture (which would prevent memorization) and batch normalization. In Table \ref{tab:results} is shown that for these settings, the model is not significantly different than the mean baseline (a p-value of $0.02$ was found, larger than our cutoff of $0.001$). Nevertheless, this means a $98\%$ probability of the model is significantly different than the mean baseline while it outperforms all other settings. Therefore, this model is chosen for future results and to be used in feature construction for the ranker.




\subsubsection{Classification}\label{sec:resultsclassification}
\begin{figure*}[t]
    \centering
    \includegraphics[width=0.48\linewidth]{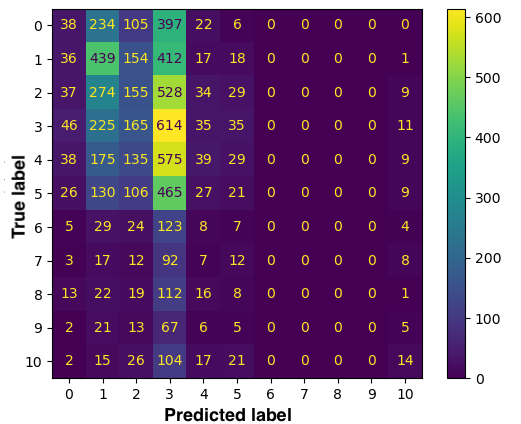}
    \includegraphics[width=0.48\linewidth]{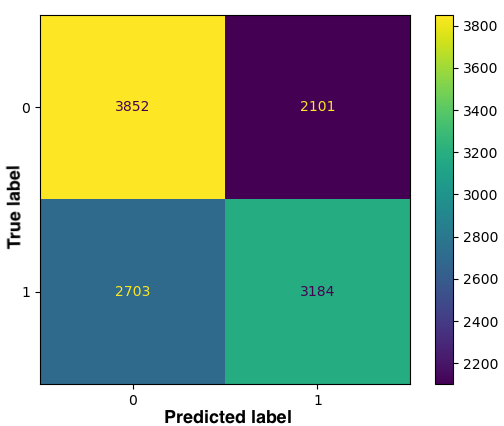}
    \caption{Confusion matrices for $11$ class (left) and $2$ class (right) data settings, using optimal parameters. Absolute numbers are shown, no normalization was applied.}
    \label{fig:confusion}
\end{figure*}

The classification was trained with identical data pre-processing settings, as described above. Similar to the regressor, we observe substantial overfitting in the training versus validation loss, despite regularizing measures. To analyze the predictions of the classifier, confusion matrices for both the $11$- and $2$-class variant were constructed, as shown in Figure \ref{fig:confusion}. 

For the $11$-class variant, we see a strong bias towards the lower engagement level labels, and moreover, some classes are not predicted at all. This is partly due to the impression filtering, in which most instances with a low impression have a high engagement level. Nevertheless, the results show that the classification model is sensitive to skewed data distribution, resulting in some classes never predicted, despite occurring quite frequently in the training set.
For the $2$-class variant, better performance is observed, with a strong trend on the confusion matrix's diagonal. As expected, the performance is not perfect, again indicating the difficulty of this task.

\subsection{Ranker}
\begin{table}[t]
\centering
    \caption{Ranker performance on test set, all differences are significant (p-value $<0.05$)}
    \begin{tabular}{|l|c|c|}
        \hline
        Model & NDCG $\uparrow$ & MRR $\uparrow$ \\\hline\hline
        With predictions & \textbf{0.620} & \textbf{0.620} \\
        Without predictions & 0.611 & 0.607\\\hline
    \end{tabular}
    \label{tab:ranker_performance}
\end{table}
Finally, we perform the ablation study on the ranker: one model was trained with the PUE from the best-performing regression model, another model was trained without those predictions. The final NDCG on the test set, as well as the MRR for both models, is shown in Table \ref{tab:ranker_performance}. The NDCG of the two models is  significantly different with a p-value of $0.0008$, indicating that using the PUE as a feature improves the results significantly, answering RQ3. Furthermore, a random, untrained model scores an NDCG of $\sim 0.47$ (results not shown), so either model improves substantially on the random baseline. Lastly, the MRR of the PUE model is also significantly better, with a p-value of $0.011$, demonstrating the effectiveness of this approach.


%% file: Sections/5-Conclusion.tex
In the upcoming field of clarifying questions in online search, we aimed to create a first ranking baseline for future work to build on. Without actual user interaction data, the task of predicting relevance to the user is a difficult one, for which perfect scores are not to be expected. Still, the results show that it is possible to predict UE from solely the lexical information in the query and CP and improve significantly on baselines, answering RQ1. Therefore we can conclude that there is evidence for user engagement in the lexical information. Nevertheless, there is a high risk of overfitting, which should be considered by future research. 

Answering RQ2, we would recommend using BERT embeddings, as they perform significantly better than TFIDF representations and are smaller, reducing the number of weights in the model. Moreover, applying the impression level filter to filter out data instances with a low number of impressions is recommended if the amount of data is sufficient for the task at hand to reduce noise and increase stability in training. Lastly, answering RQ3, the ablation study for the ranker evaluates the use of PUE as a feature in ranking the CPs. We found that using this feature yielded significant improvements in terms of NDCG and MRR compared to an identical model using only basic features.


Nevertheless, there are some limitations to the approach and assumptions in this research. The main assumption in this study is that the engagement level is a good proxy for the relevance or usefulness of a CP for the user. The MIMICS dataset that is used lacks information about the actual results that were returned for the query at hand, i.e., the Search Engine Result Page (SERP). However, it might be the case that the CP is not engaged because the relevant document was the first search result. Other reasons might be that the user did not notice the CP, or the question or answers of the CP might not be helpful. The absence of engagement with the CP does not necessarily indicate a weak match between the query and CP, but it merely indicates that the search engine performs well. Another possible option is that the CP was used to reformulate the query without directly interacting with the CP. Hence, future research might focus on incorporating features from the SERP to combine with the UE to get a better metric for CP relevance. Moreover, incorporating reformulation data might help in identifying the relevance of the CP without click interaction.

A second, more practical limitation is the fact that the presented pipeline is not yet fully automated. Future work could integrate the prediction and ranking model into one continuous pipeline. 
Furthermore, other architectures could be experimented with, such as training a custom BERT head for this task, to obtain better representations of the textual input, as explored in \cite{Rosset_suggesting_useful_questions}. 
Also, for the feature selection, other features could be constructed and considered, possibly from the SERP and reformulation data, as described above. Lastly, investigating whether we could incorporate the predicted user engagement directly in the CP generation process might be an interesting research direction.

%% file: main.bbl
\begin{thebibliography}{10}
\providecommand{\url}[1]{\texttt{#1}}
\providecommand{\urlprefix}{URL }
\providecommand{\doi}[1]{https://doi.org/#1}

\bibitem{DBLP:journals/corr/abs-2009-11352}
Aliannejadi, M., Kiseleva, J., Chuklin, A., Dalton, J., Burtsev, M.S.: Convai3:
  Generating clarifying questions for open-domain dialogue systems (clariq).
  CoRR  \textbf{abs/2009.11352} (2020), \url{https://arxiv.org/abs/2009.11352}

\bibitem{Aliannejadi_questions_in_IR}
Aliannejadi, M., Zamani, H., Crestani, F., Croft, W.B.: Asking clarifying
  questions in open-domain information-seeking conversations. In: Proceedings
  of the 42nd International ACM SIGIR Conference on Research and Development in
  Information Retrieval. p. 475–484. SIGIR'19, Association for Computing
  Machinery, New York, NY, USA (2019). \doi{10.1145/3331184.3331265},
  \url{https://doi.org/10.1145/3331184.3331265}

\bibitem{Burges_baselines}
Burges, C.J.: From ranknet to lambdarank to lambdamart: An overview. Tech. Rep.
  MSR-TR-2010-82 (June 2010),
  \url{https://www.microsoft.com/en-us/research/publication/from-ranknet-to-lambdarank-to-lambdamart-an-overview/}

\bibitem{Drosou_search_diversification}
Drosou, M., Pitoura, E.: Search result diversification. SIGMOD Rec.
  \textbf{39}(1),  41–47 (Sep 2010). \doi{10.1145/1860702.1860709},
  \url{https://doi.org/10.1145/1860702.1860709}

\bibitem{Hashemi_neural_interaction_model}
{Hashemi}, H., {Zamani}, H., {Croft}, W.B.: {Guided Transformer: Leveraging
  Multiple External Sources for Representation Learning in Conversational
  Search}. arXiv e-prints arXiv:2006.07548 (Jun 2020)

\bibitem{Kiesel_voice_query}
Kiesel, J., Bahrami, A., Stein, B., Anand, A., Hagen, M.: Toward voice query
  clarification. In: The 41st International ACM SIGIR Conference on Research \&
  Development in Information Retrieval. p. 1257–1260. SIGIR '18, Association
  for Computing Machinery, New York, NY, USA (2018).
  \doi{10.1145/3209978.3210160}, \url{https://doi.org/10.1145/3209978.3210160}

\bibitem{Krasakis_document_ranking}
Krasakis, A.M., Aliannejadi, M., Voskarides, N., Kanoulas, E.: Analysing the
  effect of clarifying questions on document ranking in conversational search.
  Proceedings of the 2020 ACM SIGIR on International Conference on Theory of
  Information Retrieval  (Sep 2020). \doi{10.1145/3409256.3409817},
  \url{http://dx.doi.org/10.1145/3409256.3409817}

\bibitem{Lehmann_user_engagement_measures}
Lehmann, J., Lalmas, M., Yom-Tov, E., Dupret, G.: Models of user engagement.
  In: Masthoff, J., Mobasher, B., Desmarais, M.C., Nkambou, R. (eds.) User
  Modeling, Adaptation, and Personalization. pp. 164--175. Springer Berlin
  Heidelberg, Berlin, Heidelberg (2012)

\bibitem{Mitra_user_interaction_autocomplete}
Mitra, B., Shokouhi, M., Radlinski, F., Hofmann, K.: On user interactions with
  query auto-completion. In: Proceedings of the 37th International ACM SIGIR
  Conference on Research \& Development in Information Retrieval. p.
  1055–1058. SIGIR '14, Association for Computing Machinery, New York, NY,
  USA (2014). \doi{10.1145/2600428.2609508},
  \url{https://doi.org/10.1145/2600428.2609508}

\bibitem{OBrien_Theoretical_UE}
O'Brien, H., Cairns, P.: Why Engagement Matters: Cross-Disciplinary
  Perspectives and Innovations on User Engagement with Digital Media. Springer
  Publishing Company, Incorporated, 1st edn. (2016)

\bibitem{Paszke_pytorch}
Paszke, A., Gross, S., Massa, F., Lerer, A., Bradbury, J., Chanan, G., Killeen,
  T., Lin, Z., Gimelshein, N., Antiga, L., Desmaison, A., Kopf, A., Yang, E.,
  DeVito, Z., Raison, M., Tejani, A., Chilamkurthy, S., Steiner, B., Fang, L.,
  Bai, J., Chintala, S.: Pytorch: An imperative style, high-performance deep
  learning library. In: Wallach, H., Larochelle, H., Beygelzimer, A.,
  d\textquotesingle Alch\'{e}-Buc, F., Fox, E., Garnett, R. (eds.) Advances in
  Neural Information Processing Systems 32, pp. 8024--8035. Curran Associates,
  Inc. (2019), \url{https://arxiv.org/abs/1912.01703}

\bibitem{Radlinski_framework}
Radlinski, F., Craswell, N.: A theoretical framework for conversational search.
  In: Proceedings of the 2017 Conference on Conference Human Information
  Interaction and Retrieval. p. 117–126. CHIIR '17, Association for Computing
  Machinery, New York, NY, USA (2017). \doi{10.1145/3020165.3020183},
  \url{https://doi.org/10.1145/3020165.3020183}

\bibitem{Reimers_sentence-bert}
Reimers, N., Gurevych, I.: Sentence-bert: Sentence embeddings using siamese
  bert-networks. In: Proceedings of the 2019 Conference on Empirical Methods in
  Natural Language Processing. Association for Computational Linguistics (11
  2019), \url{http://arxiv.org/abs/1908.10084}

\bibitem{Risch_comment}
Risch, J., Krestel, R.: Top comment or flop comment? predicting and explaining
  user engagement in online news discussions (2020)

\bibitem{Rosset_suggesting_useful_questions}
Rosset, C., Xiong, C., Song, X., Campos, D., Craswell, N., Tiwary, S., Bennett,
  P.: Leading conversational search by suggesting useful questions. In: The Web
  Conference 2020 (formerly WWW conference) (April 2020),
  \url{https://www.microsoft.com/en-us/research/publication/leading-conversational-search-by-suggesting-useful-questions/}

\bibitem{Santos_query_reformulation}
Santos, R.L., Macdonald, C., Ounis, I.: Exploiting query reformulations for web
  search result diversification. In: Proceedings of the 19th International
  Conference on World Wide Web. p. 881–890. WWW '10, Association for
  Computing Machinery, New York, NY, USA (2010). \doi{10.1145/1772690.1772780},
  \url{https://doi.org/10.1145/1772690.1772780}

\bibitem{sekulic2021user}
Sekulić, I., Aliannejadi, M., Crestani, F.: User engagement prediction for
  clarification in search. In: {Proceedings of the European Conference on
  Information Retrieval (ECIR)} (2021)

\bibitem{Straton_facebook_prediction}
{Straton}, N., {Mukkamala}, R.R., {Vatrapu}, R.: Big social data analytics for
  public health: Predicting facebook post performance using artificial neural
  networks and deep learning. In: 2017 IEEE International Congress on Big Data
  (BigData Congress). pp. 89--96 (2017)

\bibitem{Zamani_generation}
Zamani, H., Dumais, S., Craswell, N., Bennett, P., Lueck, G.: Generating
  clarifying questions for information retrieval. In: Proceedings of The Web
  Conference 2020. p. 418–428. WWW '20, Association for Computing Machinery,
  New York, NY, USA (2020). \doi{10.1145/3366423.3380126},
  \url{https://doi.org/10.1145/3366423.3380126}

\bibitem{Zamani_MIMICS}
{Zamani}, H., {Lueck}, G., {Chen}, E., {Quispe}, R., {Luu}, F., {Craswell}, N.:
  {MIMICS: A Large-Scale Data Collection for Search Clarification}. arXiv
  e-prints arXiv:2006.10174 (Jun 2020)

\bibitem{Zamani_analysis}
{Zamani}, H., {Mitra}, B., {Chen}, E., {Lueck}, G., {Diaz}, F., {Bennett},
  P.N., {Craswell}, N., {Dumais}, S.T.: {Analyzing and Learning from User
  Interactions for Search Clarification}. arXiv e-prints arXiv:2006.00166 (May
  2020)

\end{thebibliography}
